\documentstyle[preprint,eqsecnum,aps,epsfig,tighten]{revtex}

\def\be{\begin{equation}} 
\def\ee{\end{equation}} 
\def\ba{\begin{eqnarray}} 
\def\ea{\end{eqnarray}} 
\def\oo{\"{o}}

\begin{document} 
 
\title{Relativistic NN scattering without partial wave decomposition}
\author{G.\ Ramalho$^1$, A.\ Arriaga$^{1,2}$ and M.\ T.\ Pe\~na$^{3}$ \\
\vskip 2mm
{\small $^1$Centro de F{\'\i}sica Nuclear da Universidade de Lisboa, 
1649-003 Lisboa, Portugal \\
        $^2$Departamento de F{\'\i}sica, Faculdade de Ci\^encias da 
Universidade de Lisboa, 
1700 Lisboa, Portugal \\
        $^3$Centro de F{\'\i}sica das Interac\c{c}\~oes 
Fundamentais and 
Department of Physics, Instituto Superior T\'ecnico, 
Av.\ Rovisco Pais 
1049-001 Lisboa, Portugal}\\[2mm]}
\date{\today}
\maketitle

\begin{abstract}
We consider the covariant Spectator
equation with a One Boson Exchange (OBE) kernel, 
and apply it to the NN system. Relativistic effects such as 
retardation and negative-energy state components are included
in that equation. We develop a method, based on the 
Pad\'e method, to solve the Spectator equation without 
partial wave decomposition. The convergence of the partial wave decomposition 
series is tested as a function of the energy. 
The on- and off-mass-shell amplitudes are calculated.
The NN interaction was fitted to the differential cross section for NN
scattering, in the energy range below the pion production threshold.
\end{abstract}


\section{Introduction}
\label{sec1}

Electron scattering experiments off light nuclei at
momentum transfer in the few GeV/c region will provide new information on the 
intermediate and short range behavior of the nuclear interaction
\cite {CLASS03,Orden2}. In this regime the interpretation 
of data requires a description
based on relativistic kinematics and dynamics, and therefore sets 
an important challenge to theory.
It also raises a technical
computational difficulty, since
traditional numerical methods to
solve dynamical equations are based upon partial wave decomposition
(PWD) of
the NN interaction.
In fact,  
scattering and bound states are often described by partial
waves series. This procedure allows to explore nuclear symmetries and
probes expansions in powers of energy/momentum. Moreover, by reducing
the dimensionality of the
integral equations it saves numerical computational effort,
since for low energies ($<200$ MeV) the series
can be truncated to a few terms.
However, for higher energies  
a large number of partial waves
is required and the method becomes unfeasible, specially
in applications for nuclear systems with $A>2$ \cite{Fac3}.
In fact two-body scattering amplitudes are input for the 
three-body amplitudes and, 
as it is shown in Ref.\  \cite{Fac2}, above 300 MeV 
the convergence of the series requires more than 
15 partial waves.
Fortunately, present day computational limitations are not a serious objection
anymore to obtaining  directly the solution of three-dimensional integral 
equations without partial wave series.

In short, methods not based on
PWD generate numerical problems
which are i) straightforward, presenting much less 
analytical and algebraic complexity than alternative 
formulations with PWD,
ii) usable in the energy region
where PWD converge too slowly, and which is probed 
by the Jlab data
for electro- and photo- desintegration of the two- and three-nucleon 
systems,
iii) solvable within reasonable CPU time in present day
available computer power.

Since for the energy range mentioned in point ii) relativity becomes also
an issue, in this work we develop 
a method to solve relativistic quasi-potential
 equations, without PWD, for the NN system, 
in particular the Spectator equation 
\cite{Gross69,Gross92} with a realistic NN interaction. 
The present paper prepares for later
applications to photo- and electro- reactions at high energies.

By construction, the Spectator quasi-potential equation incorporates
relativistic effects such as retardation and negative-energy state components
($\rho$-spin states).
The focus here is on obtaining
a working method to solve it without PWD,
proceeding directly through a three-dimensional integration.
This is the first work that combines the two aspects, 
relativity and three-dimensional
numerical methods, in a application 
to the two-nucleon system scattering problem.
While we consider here the particular case of the Spectator equation,
the method can be extended to other quasi-potential equations.
The numerical complexity of the relativistic problem is twice as
large as the one of the non-relativistic case. 
The resulting equation couples 8 channels corresponding
to the different final helicities and $\rho$-spin cases.

In Sec.\ \ref{overview} an overview of three-dimensional
methods and of quasi-potential equations is presented.
In Sec.\ \ref{sec2} the detailed presentation of the Spectator 
equation for fermions is given.
The NN interaction is also discussed, as well as the 
strong form factors required  by the stability of the solution.
In Sec.\ \ref{sec3} the method of solving the integral equation
without partial wave decomposition is explained. In Sec.\ \ref{sec4}
results for NN on-mass-shell amplitudes, 
NN differential cross section and 
NN off-mass-shell amplitudes are shown and discussed. 
Finally, Sec.\ \ref{sec5} presents the conclusions.

\section{Background}
\label{overview}

\subsection{Three-Dimensional (3D) Methods}

For energies not larger than 300 MeV, numerical methods
without partial
wave decomposition, also referred as Three Dimensional (3D) methods
\cite{Fac2}, were applied to non-relativistic NN scattering with spin
and isospin variables, using the Bonn and Argonne interactions.
The authors used 2-particle helicity states and
the role of the symmetries of the scattering amplitudes in that basis
to  advantageously reduce the size of the numerical 3D problem.

Importantly,
their results showed that in the vicinity of 300 MeV
some of those NN amplitudes
required already 16 partial waves to converge, making
the partial wave method impracticable, specially for use in 3 nucleon
calculations. Indeed, in applications to the 3 nucleon system
presented on Ref.\  \cite{Fac3} the PWD method was realized to be 
inappropriate for 3 nucleon processes at energies higher than 250 MeV.
At those energies, both two- and three-nucleon scattering amplitudes
were shown to exhibit a strong angular dependence in the forward and backward 
scattering angles,
preventing an efficient and reasonable description 
in terms of a few partial waves.
Other non-relativistic calculations without 
partial wave decomposition  for the amplitudes
of the two- and three- scalar particle systems were published 
\cite{Thomas,Elster}. 
In Ref.\  \cite{Ramalho02} relativistic effects of 
several quasi-potential equations are compared using the 3D method 
for scalar particles.
An advantage of the 3D calculation for scalar particles based
on numeric linear system methods is that it works
on a relatively large range of energies (300-800 MeV)  
with, almost, the same number of grid points \cite{Thomas,Ramalho02}.

\subsection{Quasi-Potential (QP) equations}

Over the past years different relativistic formulations have been 
investigated for application to light nuclei, which can
be classified in two major categories:
Relativistic Hamiltonian Dynamics \cite{keister,Carlson93,Forest99} 
and Relativistic 
Field Theories \cite{Gross69,BetheS,BbS,ETbiblo,Th,EH}. 
We will concentrate here
on the second ones. They are based on
covariant field theories
with effective nucleon and meson degrees of freedom.
The non-perturbative character of the
nuclear interaction definitely requires an infinite sum of meson 
exchange diagrams, 
which can be summed effectively by means of 
integral equations, as the Bethe-Salpeter 
\cite{BetheS} 4-dimensional equation.

In principle, the kernel of the Bethe-Salpeter equation should include all 
irreducible diagrams derived from a considered Lagrangian.
In practice, models inevitably determine the truncation of the kernel 
expansion and only a restrict set of appropriate meson exchange 
diagrams is then kept. 

An usual approximation of the Bethe Salpeter equation consists on restricting 
the integral kernel
to a sum of One Boson (Meson) exchange diagrams, which is often called the
Ladder Approximation. Nevertheless, the 
this   approximation may be
questioned.
In fact, the one body limit is 
not recovered when one of the particle masses goes 
to infinity \cite{Gross69,GrossBook}.
On the other hand, the crossed-box irreducible diagrams
may be important, as confirmed by
calculations within the Feynman-Schwinger formalism
\cite{Tjon96}.  
Although  restricted only to the
bound state of scalar particles, Ref.\  
\cite{PhWal}  showed that
there are some 3-dimensional integral equations
that conveniently rearrange the series of ladder and crossed-ladder 
diagrams, yielding results equivalent to solving
the BS equation beyond the ladder approximation.   
\cite{Tjon96}. 
One of those 3-dimensional integral equations is the Spectator equation.
 
The quasi-potential (QP) equations 
are integral equations that are obtained from 
the Bethe-Salpeter equation through a restriction 
on the energy variable  yielding 3-dimensional 
integral equations.

In particular, in the Spectator equation this restriction 
is motivated by important cancellations
between box and crossed-box amplitudes. 
These cancellations are such that 
their sum reduces to ladder diagrams 
of the mass pole of the heavier particle or, 
for equal interacting masses, to ladder diagrams
of the  mass poles of the two particles.
For scalar interacting particles in the one-body limit the above
cancellation is exact order by order 
\cite{Gross69,GrossBook}.
When one of the masses is 
much larger than the other, the spectator equation in the 
ladder approximation alone gives 
practically the same result as the   
full Bethe-Salpeter equation, as shown in Ref. \cite{Gross90}. 
This outcome shows the efficiency gained
in the rearrangement of the series.
Furthermore, it has been shown that for nucleons 
(with equal masses) the non-vanishing difference between the sum of 
box and crossed-box
diagrams and the spectator ladder diagrams can be
effectively represented by 
one heavier boson exchanges \cite{Ramalho99}.
The Spectator equation,
with an appropriate OBE kernel, is therefore
suitable for applications to the two- and three-nucleon
systems, as successfully done in Refs.\   
\cite{Orden2,Gross92,Stadler97,Orden}.

\section{Spectator equation for fermions} 
\label{sec2}

In the original form of the Spectator equation
\cite{Gross69} 
one of the two particles is considered 
on-mass-shell in the intermediate state. To satisfy the 
Pauli Principle 
a new version consisting of a set of two coupled equations 
was presented in Ref.\   \cite{Gross92}. 
Since here we evaluate the scattering amplitude 
with both particles on-mass-shell in the initial state, 
the two equations reduce to a single one. 

Following Ref.\  \cite{Gross92} the scattering equation corresponding to the 
amplitude where particle 1 is on-mass-shell
in the initial and final states is written as 
\ba
& &
{\cal M}_{\alpha^\prime \beta^\prime, \alpha \beta} (p^\prime,p;P) =
\bar {\cal V}_{\alpha^\prime \beta^\prime, \alpha \beta} 
(p^\prime,p; P)  \label{GrossCompact}\\
& &
-
\frac{1}{2}
\sum\limits_{{
\matrix{{\alpha_1,\alpha_2}\cr {\beta_1,\beta_2}\cr}} }
\int \frac{d^3 k}{(2\pi)^3} \frac{m}{E_{\bf k}}
\bar {\cal V}_{\alpha^\prime \beta^\prime, \alpha_1 \beta_1} (p^\prime,k;P)
G_{\alpha_1 \beta_1, \alpha_2,\beta_2}(k;P) 
{\cal M}_{\alpha_2 \beta_2, \alpha \beta} (k,p;P),
\nonumber
\ea
where 
\be
\bar {\cal V}_{\alpha^\prime \beta^\prime, \alpha \beta}(p^\prime,p;P) =
{\cal V}_{\alpha^\prime \beta^\prime, \alpha \beta} 
(p^\prime,p; P) 
+(-1)^I {\cal V}_{\beta^\prime \alpha^\prime,\alpha  \beta} 
(-p^\prime,p; P),
\label{Vsimetric}
\ee
is the anti-symmetrized interaction kernel. In the factor $\delta=(-1)^I$ 
$I$ is the total isospin of the 2-nucleon
system. 
In the notation used $m$ is the nucleon mass,
$p$ and $p^\prime$ are the relative initial and final 4-momenta, 
$P$ the total 4-momentum. Also, the indices that represent
the Dirac components of particles 
1 and 2 are 
respectively $\alpha^\prime$ and $\beta^\prime$  for the final state,  and
$\alpha$ and $\beta$ for the initial state.
$E_{\bf k}$ is the on-mass-shell energy corresponding 
to the three momentum ${\bf k}$: 
$E_{\bf k}=\sqrt{m^2+{\bf k}^2}$.  
We considered the c.m.\  reference frame where $P=(W,0)$
($W$ is the total energy).
When both particles are on-mass-shell the relative 3-momentum 
is represented by $\overline{\bf p}$ and then $W= 2 E_{\overline{\bf p}}$.

In Eq.\ (\ref{GrossCompact}) $G_{\alpha_1 \beta_1, \alpha_2 \beta_2}(k;P)$
is the 2-nucleon Spectator propagator given by
\be
G_{\alpha_1 \beta_1, \alpha_2 \beta_2}(k;P)=
\Lambda_{\alpha_1 \alpha_2} 
\left(P/2+k \right) G_{\beta_1 \beta_2}\left(P/2-k \right),
\label{eqG2part}
\ee
where $\Lambda_{\alpha_1 \alpha_2}
(k_1)$ is the particle 1 positive-energy projector and 
$G_{\beta_1 \beta_2}(k_2)$ the particle 2 Dirac propagator.
In Eq.\  (\ref{GrossCompact}) the asymmetry due to the propagator $G$ 
is only apparent, since the anti-symmetrization is implemented by means 
of the kernel in Eq. (\ref{Vsimetric}).

To write the scattering amplitude  
in terms of the helicity and $\rho$-spin 
we perform the change of basis similar to the one of the
Ref.\  \cite{Gross92}:
\ba
& &
{\cal M}_{\lambda_1^\prime \lambda_2^\prime, \lambda_1 \lambda_2}^{
\rho_1^\prime \rho_2^\prime, \rho_1 \rho_2}(p^\prime,p;P)= 
\label{Mlambda} \\
& &
\sum\limits_{\footnotesize{
\matrix{{\alpha^\prime, \beta^\prime}\cr {\alpha, \beta}\cr}} }
\bar u_{1\; \alpha^\prime}^{\rho_1^\prime}(p^\prime, \lambda_1^\prime) 
\bar u_{2\; \beta^\prime}^{\rho_2^\prime}(p^\prime,\lambda_2^\prime) 
{\cal M}_{\alpha^\prime \beta^\prime, \alpha \beta} (p^\prime,p;P) 
u^{\rho_1}_{1\; \alpha}(p,\lambda_1)
u^{\rho_2}_{2\; \beta}(p,\lambda_2),
\nonumber
\ea
where $u_j^{\pm}$ with  $j=1,2$ are the asymptotic states 
of the particles (see Appendix \ref{apHelicidades}),
$\lambda_1$ ($\lambda_1^\prime$) and $\lambda_2$ ($\lambda_2^\prime$) are the initial
(final) helicities 
respectively for particles 1 and 2. 
The indices $\rho_j$ and $\rho_j^\prime$ with $j=1,2$ express 
the initial and final $\rho$-spin states for particle $j$. 
A similar expression holds for ${\cal V}$.

\begin{figure}
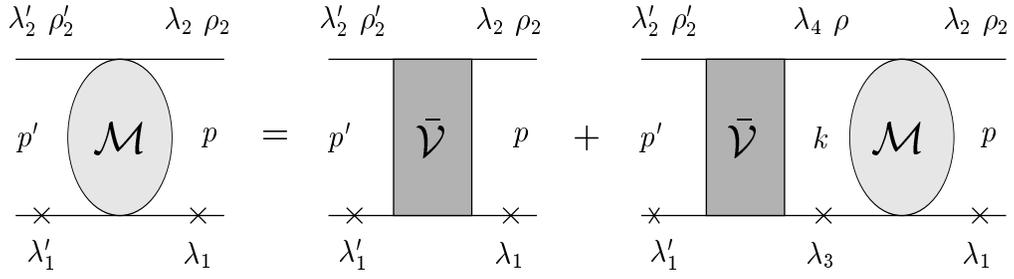

\caption{Helicity representation of the Spectator equation.}
\label{CompactGross}
\end{figure}

The scattering amplitude 
in terms of the helicity and $\rho$-spin reads:
\ba
& &
{\cal M}_{\lambda_1^\prime \lambda_2^\prime, \lambda_1 \lambda_2}^{
+ \rho_2^\prime, + \rho_2}(p^\prime,p;P)=
\bar 
{\cal V}_{\lambda_1^\prime \lambda_2^\prime, \lambda_1 \lambda_2}^{
+ \rho_2^\prime, + \rho_2}(p^\prime,p;P) 
\label{EqMlambda}\\
& &
-\sum\limits_{{\rho, \lambda_3, \lambda_4}}
\int \frac{d^3 k}{(2\pi)^3} 
\bar {\cal V}_{\lambda_1^\prime \lambda_2^\prime, 
\lambda_3 \lambda_4}^{+ \rho_2^\prime, + \rho}(p^\prime,k;P)
g^\rho (k;W)
{\cal M}_{\lambda_3 \lambda_4, \lambda_1 \lambda_2}^{
+ \rho, + \rho_2}(k,p;P),
\nonumber
\ea
where
\ba
& &
g^+(k;W)=\frac{1}{2}
\left(\frac{m}{E_{\bf k}}\right)^2 
\frac{1}{2 E_{\bf k}-W-i\varepsilon} 
\label{eqg+}\\ 
& &
g^-(k;W)=-\frac{1}{2}
\left(\frac{m}{E_{\bf k}}\right)^2 \frac{1}{W}.
\label{eqg-}
\ea
Note that in the Spectator formalism 
one of the particles is always on-mass-shell 
in a positive-energy state and 
that is why in Eq.\  (\ref{EqMlambda}) there is always 
a positive  $\rho$-spin state in the initial, 
intermediate and final states.
The diagrammatic representation of the Eq.\  (\ref{EqMlambda}) 
is given in the Fig.\  \ref{CompactGross}.
The crosses on the lines mean that the corresponding particles are 
on mass-shell with positive energy.

\begin{figure}
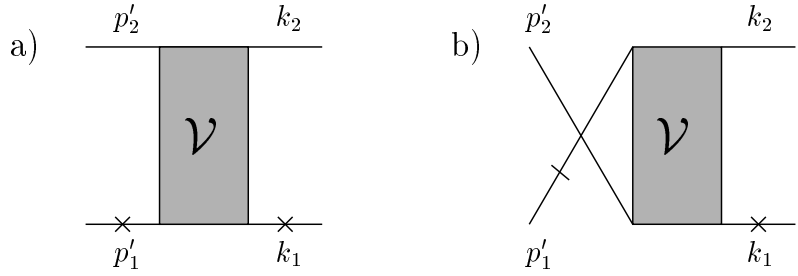

\caption{a) Direct term of the kernel 
${\bar {\cal V}}$.
b) Exchange term of the kernel ${\bar {\cal V}}$.}
\label{figVsim}
\end{figure}

The kernel $\bar {\cal V}$ contains two terms, 
the direct term and the 
exchange term (see Eq.\  (\ref{Vsimetric})),
which are represented in Fig.\  \ref{figVsim}.  
In this work we use an OBE kernel including $\pi$, $\sigma$, 
$\rho$ and $\omega$ meson exchanges. 
For $\pi$ exchange we consider a mixture of 
pseudo-scalar (PS) and pseudo-vector (PV) couplings, defined as:
\be
\Lambda_{\pi}(p^\prime,k)=\lambda_\pi \gamma^5+(1-\lambda_\pi)
\frac{(\not p^\prime -\not k)}{2m} \gamma^5,
\ee 
where $0 \le \lambda_\pi \le 1$ is the admixture parameter.

The kernel parameters are presented in table \ref{tabNN}
along with the cut-off parameters of the form factors 
(see Sec.\  \ref{secStrongFF}).
The $m_\pi$, $m_\rho$ and $m_\omega$ parameters are fixed 
by the experimental values for physical mesons. 
The  remaining 11 parameters were
fitted to NN scattering data.  
We may anticipate at this point
that the best results obtained favor the PV coupling ($\lambda_\pi=0$).
The model used for the kernel is very similar to the one 
of the Refs.\  \cite{Gross92,Stadler97}.
Explicit expressions for ${\cal V}$ can be found 
in Appendix \ref{apKernel}.

\subsection{Prescription for the Exchange Kernel}

For the direct term the 4-momentum transfer $q$ is given by
\be
q^2=(p_1^\prime-k_1)^2
\ee
while for the exchange term by
\be
{\widehat q}^2=(p_1^\prime-k_2)^2.
\label{eqQtrans}
\ee

The relation of these four-momenta with the relative 
four-momenta is given by
\ba
& &p_1^\prime =(E_{\bf p^\prime},{\bf p^\prime}) \\
& &p_2^\prime =(W-E_{\bf p^\prime},-{\bf p^\prime}) \\
& &k_1 =(E_{\bf k},{\bf k}) \\
& &k_2 =(W-E_{\bf k},-{\bf k}).
\ea

When particle 1 is on-mass-shell with
positive-energy the momentum transfer reads:
\ba
& &-q^2=({\bf p}^\prime -{\bf k})^2-(E_{{\bf p}^\prime}-E_{\bf k})^2 
\label{q2direct}\\
& &-{\widehat q}^2=({\bf p}^\prime +{\bf k})^2-(E_{{\bf p}^\prime}+E_{\bf k}-W)^2,
\label{q2exchange}
\ea
respectively for the direct and exchange terms of the kernel
(see Fig.\  \ref{figVsim}).
We may then conclude that, while the direct term has no 
singularities in the meson propagators (we have always 
$\mu^2 -q^2 > 0$),
the exchange term may have the singularity corresponding to an 
on-mass-shell exchanged meson ($\mu^2={\widehat q}^2$). 
The condition $\mu^2={\widehat q}^2$ indicates 
a real meson production from off-mass-shell 
nucleon states. However, since a real meson production process 
is not allowed below the pion production threshold,
the singularity of $\widehat {\cal V}$ has no physical correspondence. 
It is simply a consequence of the anti-symmetrization of the kernel 
(see Eq.(\ref{Vsimetric})). As shown by Gross {\it et al} this spurious
singularity is cancelled by higher order diagrams \cite{Gross92}.

In the numerical applications of the Spectator equations
two prescriptions were considered so far to deal with the
spurious singularities \cite{Gross92}:
\begin{itemize}
\item 
the "principal part prescription": the singularity is included 
but only the Principal Part of the integral is kept. This corresponds 
to the replacement 
$$
\int \frac{1}{\mu^2 -\widehat q^2-i\varepsilon} \to
{\cal P}\int \frac{1}{\mu^2 -\widehat q^2}=
\int \left(\frac{1}{\mu^2 -\widehat q^2-i\varepsilon}-i \pi 
\delta(\mu^2 -\widehat q^2) \right).
$$ 
\item 
the "energy independent prescription": the momentum transfer is 
modified in order to remove the singularity. This amounts to the
replacement 
$${\widehat q}=p_1^\prime-k_2 \to q= p_1^\prime-k_1,$$ 
\end{itemize}
that is, the momentum transfer is the same for the propagator in both 
direct and exchange diagrams, leaving the meson propagator denominator
independent of the two-nucleon system energy $W$ 
(see Eq.\  (\ref{q2direct})).

We now introduce the on-shell prescription, 
an alternative adopted by us in the present work.

Eq.\  (\ref{q2exchange}) can be rewritten as follows:
\be
-\widehat q^2=({\bf p}^\prime+{\bf k})^2-
(E_{{\bf p}^\prime}-E_{\bf k})^2-
(W-2 E_{{\bf p}^\prime})(W- 2E_{\bf k}).
\ee
The last term of this equation generates 
the spurious singularities mentioned above, 
and vanishes when both particles are on-mass-shell 
either in the initial state ($W=2 E_{\bf k}$) or in the 
final state ($W=2 E_{{\bf p}^\prime}$).
We therefore define the ``on-mass-shell prescription'', 
by taking 
for $-\widehat q^2$ the expression
\be
-\widehat q^2 \to ({\bf p}^\prime+{\bf k})^2-
(E_{{\bf p}^\prime}-E_{\bf k})^2.
\label{q2exchange2}
\ee
With this choice the exchange kernel $\widehat {\cal V}$ 
is consistent with the Feynman rules in 
the on-mass-shell limit. This is a physical 
argument in favor of the proposed prescription.
Furthermore, the plus sign in 
Eq.\  (\ref{q2exchange2}) implies that 
when the direct term in a particular interaction is dominant
in the forward direction, then 
the exchange term dominates 
in the backward direction, as it happens 
for interactions mediated by scalars.
We point out that this property is not
satisfied by the "energy independent prescription", where 
the exchange kernel $\widehat q$ is replaced by the 
direct kernel momentum $q$.
Note also that the "on-mass-shell prescription" is energy independent.

\subsection{Properties and Symmetries of the Amplitudes}

Using parity, time reversal and particle interchange symmetries 
we can decrease the number of independent amplitudes 
\cite{Bystricky78,BrownJackson}.
In particular, parity invariance 
reduces the 16 helicity amplitudes
with $\rho^\prime=+1$, $\rho=+1$ states to 8 independent ones, according to:
\ba
& M_1&=
{\cal M}_{++,++}
(\mbox{p}^\prime,u; \mbox{p},1)=
{\cal M}_{--,--} 
(\mbox{p}^\prime,u; \mbox{p},1) 
\label{eqM1}  \\
& M_2&=
{\cal M}_{--,++}
(\mbox{p}^\prime,u; \mbox{p},1)=
{\cal M}_{++,--}
(\mbox{p}^\prime,u; \mbox{p},1) \\
& M_3&=
{\cal M}_{+-,+-}
(\mbox{p}^\prime,u; \mbox{p},1)=
{\cal M}_{-+,-+}
(\mbox{p}^\prime,u; \mbox{p},1) \\
& M_4&=
{\cal M}_{-+,+-}
(\mbox{p}^\prime,u; \mbox{p},1)=
{\cal M}_{+-,-+}
(\mbox{p}^\prime,u; \mbox{p},1) \\
&M_5&=
{\cal M}_{-+,++}
(\mbox{p}^\prime,u; \mbox{p},1)=
-{\cal M}_{+-,--}
(\mbox{p}^\prime,u; \mbox{p},1) \\
&M_6&=
{\cal M}_{+-,++}
(\mbox{p}^\prime,u; \mbox{p},1)=
-{\cal M}_{-+,--} 
(\mbox{p}^\prime,u; \mbox{p},1) \\
&M_7&=
{\cal M}_{++,+-}
(\mbox{p}^\prime,u; \mbox{p},1)=
-{\cal M}_{--,-+}
(\mbox{p}^\prime,u; \mbox{p},1) \\
&M_8&=
{\cal M}_{--,+-}
(\mbox{p}^\prime,u; \mbox{p},1)=
-{\cal M}_{++,-+}
(\mbox{p}^\prime,u; \mbox{p},1). 
\label{eqM8}
\ea
In the case $\rho^\prime=-1$ the right hand side should  
include the phase 
$(-1)^{\frac{1-\rho^\prime}{2}}=-1$.

The above relations are valid either $\mbox{p}^\prime$ is on- or 
off-mass-shell. Restricting to the on-mass-shell situation, further
relations arise:
\ba
& &M_7=M_6 \\
& &M_8=-M_5 \\
& &M_5=-M_6.
\label{simetriaM5M6}
\ea
The last identity is valid only for identical particles 
(exact isospin symmetry).
As a result we are left with only 5 independent amplitudes.
We point out that these relations are independent of the interaction used 
and of any prescription adopted, and are merely a 
result of the symmetries mentioned above.
The relations (\ref{eqM1})-(\ref{eqM8}) were tested 
numerically as a check to the code, since we did not explicitly
impose the symmetries to reduce the number of equations.

If we consider instead a QP equation of the instantaneous type
(no retardation), e.g. the Blanklenbecler-Sugar \cite{BbS}
 or the Equal-Time \cite{ETbiblo} equations, where the particles have always 
the same energy, we could use more symmetry properties 
to reduce further the number of off-mass-shell amplitudes.

Finally we note that, for equations of the 
instantaneous type, 
a convenient combination of the helicity states
defines states of well defined parity and  two-body spin and helicity.
The use of that basis states reduces the size of the numerical 
problem by block-diagonalizing
the set of equations. 
This was done for example in Ref.\  
\cite{Fac3,Fac2} in a non-relativistic framework 
(no $\rho^\prime=-1$ states).
Nevertheless, for 3-body applications \cite{Fac3} 
the helicity combination has to be inverted by calculating back the amplitudes
in the asymptotic basis of uncoupled helicities.

\subsection{Strong Form Factors}
\label{secStrongFF}

Nucleons are not elementary particles and their hadronic structure 
has to be taken into account through form factors.
Mathematically, such form factors provide the necessary regularization
of the integrals for the high order loops.
Since in this work we solve the scattering equation 
without partial wave expansion,
a careful study of the integrand function had to be 
performed in order to determine form factor functions adequate for 
the convergence of the method used.

The starting point for the choice of form factors was 
the decomposition suggested by Riska and Gross \cite{Riska87}
\be
F_i(p_j^\prime,k_j)= 
f_{m_i}(q^2) f_N(p_j^{\prime\; 2}) f_N(k_j^2),
\label{eqFm}
\ee
where $p_j^\prime$ ($k_j$) is the final (initial) 
momentum of the nucleon $j$,  
$q=p_j^\prime-k_j$ is the transfer momentum, $f_{m_i}$ is the 
form factor of meson $i$ and $f_N$ the nucleon form factor.
This factorization is adequate 
to describe the electromagnetic interaction with nucleons, and the
scalar functions $f_N$ and $f_m$ could be interpreted as strong interaction 
vertex corrections and self-energy contributions of nucleons and mesons. 
As shown in Ref.\  \cite{Ramalho99}, however, some care has to be taken in the
functional forms of the form factors considered in the above 
prescription. In fact, the most used momentum dependence 
of $f_N$ \cite{Gross92,Stadler97} 
induces additional spurious singularities
in a 4-dimensional framework, which can be eliminated 
by distorting conveniently the contour for energy integration 
around them \cite{GrossPrivate}.
These spurious singularities, however, do not occur  
within a QP context.

For 
the meson form factor our choice is 
\be
f_{m_i}(q^2)= \frac{\Lambda_{m_i}^2}{\Lambda_{m_i}^2-q^2},
\label{eqfm}
\ee
where $\Lambda_{m_i}$ is the cut off of meson $i$.
Note however, that 
by construction the functions $f_{m_i}(q^2)$ only modify the kernel 
for large $q^2$ (for $q^2=0$ we have $f_{m_i}(q^2)=1$),
and do not suppress the large momentum dependence of 
${\cal V}(k,k;P)$.

For the $f_N$ form factors we take
\be
f_N(k_j^2)=\left[ \frac{\tilde \Lambda_N^2}
{\tilde \Lambda_N^2+(m^2-k_j^2)^2} \right]^n,
\label{eqfn}
\ee
with 
\be
\tilde \Lambda_N^2= \Lambda_N^2-m^2.
\ee
being $\Lambda_N$ the nucleon cut-off. 
The $f_N$ functions 
regulate the asymptotic behavior
of both ${\cal V}(k,k;P)$ and ${\cal V}(k,p;P)$.
The factorization (\ref{eqFm}) has been applied 
by Gross and collaborators in several applications 
of the Spectator equation \cite{Gross92,Stadler97}. 
The function (\ref{eqfn}) with $n=1$ has been used for the 
first time in Ref.\  \cite{Stadler97} and 
kept in Spectator equation applications since then. 
The meson form factor used in the same applications differs 
from (\ref{eqfm}) for large $q^2$. 
The factorization (\ref{eqFm}) with 
the presented $f_m$ and $f_N$ ($n=1$) functional forms has been 
also used in Ref.\  \cite{Caia02}, but in 
an instantaneous quasi-potential framework
(where $k=(0, {\bf k})$). 
In the present calculation we found that in order to solve the
Spectator equation with 3D methods the value $n=1$,
used in previous applications with PWD,
was not large enough to allow the convergence 
of ${\cal M}(k,p;P)$ for large $|{\bf k}|$. We had to take instead
$n=2$.
This is mostly due to 
the meson propagator behavior 
which peaks for forward and backward scattering angles 
at  high values of the momentum \cite{Ramalho03a}.
Using the PWD method the meson propagator peak is smeared
by the angle integration, but in the 3D method the structure of the
propagator cannot be smoothened.

\section{Solution of the Integral Equation without 
Partial Wave Decomposition}
\label{sec3}

In order to solve the scattering equation 
analytical or numerically we need to specify 
the scattering conditions, that is the 
initial and final momenta. 
We choose a reference frame where the incoming momentum 
is along the $z$ axis and 
express the initial, final and intermediate 
momenta in terms of spherical coordinates 
\ba
& &{\bf p}=(\mbox{p},0,0) 
\label{eqCoor1} \\
& &{\bf p^\prime}=(\mbox{p}^\prime, \theta^\prime, \varphi^\prime)
\label{eqCoor2} \\
& &{\bf k}=(\mbox{k}, \theta, \varphi).
\label{eqCoor3}
\ea
where $\mbox{p}^\prime=|{\bf p^\prime}|$, $\mbox{p}=|{\bf p}|$ 
and $\mbox{k}=|{\bf k}|$.

\subsection{Integration of the Azimuthal Angle}

To perform the $\varphi$ integration 
we need to apply on the scattering
amplitude a rotation 
of an angle $\varphi$ around the $z$ axis. 
In Appendix \ref{apAmpRot} we show that:
\be
{\cal M}_{\lambda_3 \lambda_4, 
\lambda_1 \lambda_2}^{+\; 
\rho_2^\prime,+ \; \rho_2}
(\mbox{k},\theta, \varphi; \mbox{p};W)=
e^{i(\lambda_1- \lambda_2)\frac{\varphi}{2}}
{\cal M}_{\lambda_3 \lambda_4, \lambda_1 \lambda_2}^{
+\; \rho_2^\prime, +\; \rho_2}(\mbox{k},\theta,0;\mbox{p};W),
\label{eqInvariante}
\ee
where the ${\bf p}$ angles are omitted by simplicity.

Inserting Eq.\ (\ref{eqInvariante}) in the 
Spectator Eq.\ (\ref{EqMlambda}), and taking $\varphi^\prime=0$,  
we can factorize the $\varphi$ dependence obtaining
\ba
& &
{\cal M}_{\lambda_1^\prime \lambda_2^\prime, \lambda_1 \lambda_2}^{
+ \rho_2^\prime, + \rho_2}
(\mbox{p}^\prime,\theta^\prime,0;\mbox{p};W)=
\bar 
{\cal V}_{\lambda_1^\prime \lambda_2^\prime, \lambda_1 \lambda_2}^{
+ \rho_2^\prime, + \rho_2}
(\mbox{p}^\prime,\theta^\prime,0;\mbox{p};W) 
\label{EqMlambda2}\\  
& &
-
\sum\limits_{{\rho, \lambda_3, \lambda_4}}
\int \frac{\mbox{k}^2 d\mbox{k}}{2 \pi}
\int \frac{d\cos \theta}{2 \pi}  V_{\lambda_1^\prime \lambda_2^\prime, 
\lambda_3 \lambda_4}^{+ \rho_2^\prime, + \rho}
(\mbox{p}^\prime,\theta^\prime,\mbox{k},\theta;\bar \lambda,W) 
g^\rho (\mbox{k};W) {\cal M}_{\lambda_3 \lambda_4, 
\lambda_1 \lambda_2}^{+ \rho, + \rho_2}
(\mbox{k},\theta,0;\mbox{p};W),
\nonumber
\ea
where 
\be
V_{\lambda_1^\prime \lambda_2^\prime, 
\lambda_3 \lambda_4}^{+ \rho_2^\prime, + \rho}
(\mbox{p}^\prime,\theta^\prime,\mbox{k},\theta;\bar \lambda,W)= 
\frac{1}{2\pi}
\int d \varphi \; 
e^{i \bar \lambda \varphi}
\bar {\cal V}_{\lambda_1^\prime \lambda_2^\prime, 
\lambda_3 \lambda_4}^{+ \rho_2^\prime, + \rho}
(\mbox{p}^\prime,\theta^\prime,\varphi^\prime;
\mbox{k},\theta,\varphi;W),
\label{defV}
\ee
and
\be
\bar \lambda= \frac{\lambda_1-\lambda_2}{2}
\ee
can take the values $0,\pm 1$. 

The scattering equation, Eq.\  (\ref{EqMlambda2}), 
is now in 2-dimensional form and includes the 
propagator function $g^\rho(\mbox{k};W)$ which has 
a real pole at $\mbox{k}=\overline{\mbox{p}}$ 
($W=2E_{\bf k}$) for $\rho=+1$. 
Performing the contour integration we obtain
\ba
& &
{\cal M}_{\lambda_1^\prime \lambda_2^\prime, \lambda_1 \lambda_2}^{
+ \rho_2^\prime, + \rho_2}(\mbox{p}^\prime,\theta^\prime,0; \mbox{p};W)=
\bar 
{\cal V}_{\lambda_1^\prime \lambda_2^\prime, \lambda_1 \lambda_2}^{
+ \rho_2^\prime, + \rho_2}(\mbox{p}^\prime,\theta^\prime,0;\mbox{p};W) 
\nonumber \\  
& &
-
\sum\limits_{{\rho, \lambda_3, \lambda_4}}
{\cal P}\int \frac{\mbox{k}^2 d\mbox{k}}{2\pi} 
\int \frac{d\cos \theta}{2\pi}  V_{\lambda_1^\prime \lambda_2^\prime, 
\lambda_3 \lambda_4}^{+ \rho_2^\prime, + \rho}
(\mbox{p}^\prime,\theta^\prime;\mbox{k},\theta;\bar \lambda, W) 
g^\rho_{\varepsilon=0} (\mbox{k};W) {\cal M}_{\lambda_3 \lambda_4, 
\lambda_1 \lambda_2}^{+ \rho, + \rho_2}
(\mbox{k},\theta,0;\mbox{p};W) \nonumber \\
& &
-i
\frac{m^2 \overline{\mbox{p}}}{4W}
\sum\limits_{{\lambda_3, \lambda_4}}
\int \frac{d\cos \theta}{2\pi}  
V_{\lambda_1^\prime \lambda_2^\prime, 
\lambda_3 \lambda_4}^{+ \rho_2^\prime, + \,+}
(\mbox{p}^\prime,\theta^\prime;\overline{\mbox{p}},\theta; \bar \lambda, W) 
{\cal M}_{\lambda_3 \lambda_4, 
\lambda_1 \lambda_2}^{+ \,+, + \rho_2}
(\overline{\mbox{p}},\theta,0;\mbox{p};W).
\label{EqMlambda3}
\ea
In this equation we use $g^\rho_{\varepsilon=0} (\mbox{k};W)$ to represent 
the $\varepsilon=0$ limit of $g^\rho (\mbox{k};W)$ 
(see Eqs.\ (\ref{eqg+})-(\ref{eqg-})).
The multiplicative factor of the last term is a result of the residue
\be
\frac{1}{2}
\left[
\frac{1}{2}
\left(\frac{m}{E_{\bf k}} \right)^2
\frac{\mbox{k}^2}
{\left| \frac{\mbox{d}}{\mbox{dk}} (2 E_{\bf k}-W)\right|}
\right]_{\mbox{k}=\overline{\mbox{p}}} =
\frac{m^2 \overline{\mbox{p}}}{4W}. 
\ee
In Eq.\  (\ref{EqMlambda3}) only the function $g^+(\mbox{k};W)$ 
has a singularity but we include the 
principal part symbol ${\cal P}$ also for $\rho=-1$ 
for the sake of simplicity.

The $\varphi$-integration (\ref{defV}) can 
be performed either analytical or 
numerically. In Refs.\  
\cite{Fac3,Fac2,Thomas,Elster,Ramalho02} the numerical
integration was made. Here we choose the  analytical integration  
in order to minimize computing time. 
The complexity of the analytical expressions 
depends crucially upon the $f_m$ form factors. 
Relatively simple results are obtained 
for the monopolar choice (\ref{eqfm}).
In this case  we can write Eq.\  (\ref{defV}) as
\be
\int d \varphi 
e^{i \bar \lambda \varphi}
{\cal V}_{\lambda_1^\prime \lambda_2^\prime, \lambda_3 \lambda_4}^{
+ \rho_2^\prime, + \rho_2}(p^\prime,p; P)=
\sum\limits_{\Gamma} C_{\Gamma}
\int d \varphi \frac{e^{i \Gamma \varphi}}
{(a-b \cos \varphi)(c-b \cos \varphi)^2},
\ee
where $\Gamma=0,\pm 1,\pm 2$, and $C_\Gamma$ is a known function
of the momentum magnitudes and polar angle.  
Any term of the last equation is subsequently easily integrated over.
Meanwhile, in Ref.\  \cite{Caia}, where the equation for $\pi$-N scattering
is solved also without PWD, the authors came also to the same analytical
technique but for an integrand not including form factors.
Details of the analytical structure of the kernel can be found in 
Appendix \ref {apKernel} and the main steps for the analytical integration 
are given in Appendix \ref{apIntPhi}.

\subsection{Numerical Method: Pad\'e Method}
\label{secPade}

The relevant variables in the scattering equation (\ref{EqMlambda3})
may be emphasized by the following convenient change of notation: 
we denote each of the eight possible 
helicity and $\rho$-spin combinations by a single index  
$I^\prime=\left\{ \rho_2^\prime, \lambda_1^\prime,\lambda_2^\prime \right\}$, 
$I_k=\left\{ \rho, \lambda_3,\lambda_4 \right\}$ and 
$I_0=\left\{ \rho_2, \lambda_1,\lambda_2 \right\}$,
according to table  \ref{tabelaMI}
($I^\prime$, $I_k$ and $I_0$ have nothing to do
with the total isospin $I$); we omit the total energy $W$, 
incoming momentum $\mbox{p}$ and total isospin $I$ dependences
and perform the following  substitutions
\ba
{\cal M}_{\lambda_1^\prime \lambda_2^\prime, \lambda_1 \lambda_2}^{
+ \rho_2^\prime, + \rho_2}(\mbox{p}^\prime,\theta^\prime,0;\mbox{p};W) & \to &
{\cal M}_{I^\prime,I_0}(\mbox{p}^\prime,u) \nonumber \\
\bar {\cal V}_{\lambda_1^\prime \lambda_2^\prime, \lambda_1 \lambda_2}^{
+ \rho_2^\prime, + \rho_2}(\mbox{p}^\prime,\theta^\prime,0;\mbox{p};W) & \to &
\bar {\cal V}_{I^\prime,I_0}(\mbox{p}^\prime,u) \nonumber \\
V_{\lambda_1^\prime \lambda_2^\prime, 
\lambda_3 \lambda_4}^{+ \rho_2^\prime, + \rho}
(\mbox{p}^\prime,\theta^\prime;\mbox{k},\theta;\bar \lambda ,W)  & \to &
V_{I^\prime,I_k}(\mbox{p}^\prime,u; \mbox{k},v), \nonumber
\ea
where 
\ba
&& u=\cos \theta^\prime \\
&& v=\cos \theta.
\ea

Finally Eq.\  (\ref{EqMlambda3}) simplifies to
\ba
& &
{\cal M}_{I^\prime,I_0}(\mbox{p}^\prime,u)=
\bar {\cal V}_{I^\prime,I_0}(\mbox{p}^\prime,u) \nonumber \\
& &
-\sum\limits_{I_k=1}^8 {\cal P} \int_0^\infty
\frac{\mbox{k}^2 \mbox{dk}}{2\pi} \int_{-1}^1
\frac{dv}{2\pi} V_{I^\prime,I_k}(\mbox{p}^\prime,u;\mbox{k},v)
g^{I_k}(\mbox{k};W){\cal M}_{I_k,I_0}(\mbox{k},v) 
\label{EqMlambda4}
\\
& &
-i \frac{m^2 \overline{\mbox{p}}}{4W} 
\sum\limits_{I_k=1}^4 
 \int_{-1}^1\frac{dv}{2\pi} V_{I^\prime,I_k}
(\mbox{p}^\prime,u;\overline{\mbox{p}},v)
{\cal M}_{I_k,I_0}(\overline{\mbox{p}},v),
\nonumber 
\ea
with
\be
g^{I_k}(\mbox{k};W)= 
\left\{\matrix{ g^+_{\varepsilon=0}(\mbox{k}; W)& & \mbox{if} & & I_k=1,..,4\cr
g^-(\mbox{k}; W) & & \mbox{if} & & I_k=5,..,8\cr}
\right..
\ee

In order to obtain a numerical solution of Eq.\  (\ref{EqMlambda4}) 
we carry out a discretization of the integral variables, 
$\mbox{k} \in [0,+\infty[$ and $v \in [-1,1]$, and use a gaussian quadrature 
integration technique. 
We choose a grid of $N_p+1$ points for all momenta, 
$\mbox{p}^\prime$, $\mbox{k}$ and $\mbox{p}$, 
and a grid of $N_u+1$ points for the angular variables $u$ and $v$.
With this discretization procedure we transform the integral 
equation (\ref{EqMlambda3}) into an algebraic set of equations:
\be
M=V+C \cdot M,
\label{eqSys1}
\ee
where $M$ and $V$ are the matrix vectors 
${\cal M}_{I^\prime, I_0} (k_{i^\prime},v_{j^\prime})$ and
${\cal V}_{I^\prime, I_0} (k_{i^\prime},v_{j^\prime})$.
The $C$ matrix can be decomposed as 
$C=A+B$. For $I_k=1,..,4$, we have
\ba
A_{(I^\prime i^\prime j^\prime),(I_k i j)}&=&
-\frac{w_i^\prime  h_j}{(2\pi)^2}
k_i^2 V_{I^\prime,I_k}(k_{i^\prime},u_{j^\prime};
k_i,u_j) g^+(k_i;W) \\
B_{(I^\prime i^\prime j^\prime),(I_k i j)}&=&-
\frac{h_j}{(2\pi)^2}
\frac{m^2\overline{\mbox{p}}}{2} 
\left(
i \frac{\pi}{W} -\Delta S\right)
V_{I^\prime,I_k}(k_{i^\prime},u_{j^\prime};
\overline{\mbox{p}},u_j)  \delta_{i,N_{p}+1}
\ea
with
\ba
\Delta S &=& S-S^\prime \\
S&=& {\cal P} \int_0^{+\infty}
dk \; \frac{k}{E_{\bf k}^2} 
\frac{1}{2E_{\bf k} -W}=
-\frac{1}{W}\log \frac{W-2m}{2m} \\
S^\prime&=&
\sum\limits_{i=1}^{N_p} w_i^\prime 
\frac{k_i}{E_{k_i}^2}\frac{1}{2E_{k_i}-W}.
\ea
For $I_k=5,..,8$, we have
\ba
A_{(I^\prime i^\prime j^\prime),(I_k i j)}&=&-
\frac{w_i^\prime  h_j}{(2\pi)^2}
k_i^2 V_{I^\prime,I_k}(k_{i^\prime},u_{j^\prime};
k_i,u_j) g^-(k_i;W) \\
B_{(I^\prime i^\prime j^\prime),(I_k i j)}&=& 0.
\ea
In the previous equations $w_i^\prime$ and $h_j$ are 
respectively the gaussian weights for the variables 
$k_i$ and $u_j$. The momentum grid is obtained 
from a $x_i \in \;  ]0,1[$ grid by a change of variable
$$
k_i=\Lambda \frac{x_i}{1-x_i},
$$
where typically we take $\Lambda \simeq 0.5$ $m$.
In order to determine the on-mass-shell 
forward scattering amplitude
we add the points $k_{N_p+1}=\overline{\mbox{p}}$ 
and $u_{N_u+1}=1$ with zero weight.

The dimension of the above matrices is
$n=8(N_{p}+1)(N_u+1)$, which is
a large number when $N_p$ and $N_u$ are of the order of 20. 
Therefore, a standard matrix inversion method 
requiring a large computer memory (for double precision complex numbers), 
becomes impracticable.
To overcome this limitation we use instead the Pad\'e method, 
which gives a fast estimate of the result of the Born series for 
the coupled set of equations
\be
M=\lambda V+ \lambda C \cdot M,
\ee  
where the $\lambda$ parameter is introduced 
by convenience and set to 1 at the end of the calculation.   
The usual power expansion for $2N+1$ terms reads
\be
M=\lambda M^{(1)}+ \lambda^2 M^{(2)}+  \lambda^3 M^{(3)}+ ...
+  \lambda^{2N +1} M^{(2N+1)}. 
\label{matrizPADE}
\ee 
The $M^{(i)}$ vectors are evaluated by
\ba
M^{(1)}&=&V \\
M^{(i+1)}&=&C \cdot  M^{(i)}.
\label{TaylorRec}
\ea 
and any element $m$ of the $M$ vector given by
\be
m=\lambda m_1+ \lambda^2 m_2+  \lambda^3 m_3+ ...
+  \lambda^{2N +1} m_{2N+1},
\label{mesc} 
\ee
is approximated by a rational function of $\lambda$
\be
m_{Pade}(\lambda)
=\lambda \frac{a_0+\lambda a_1 +...+\lambda^{N}a_{N}}{
1 +\lambda b_1 +...+ \lambda^{N}b_{N}},
\label{mPadeS}
\ee
where the $2N+1$ $a_l$ and $b_l$ coefficients are 
determined through the $2N+1$ $m_l$ coefficients, after equating
Eqs.\ (\ref{mesc}) and (\ref{mPadeS}).
This method is known in the literature as SPA 
(Scalar Pad\'e Approximant) \cite{Gersten74} 
and  $m_{Pade}$ denotes the Pad\'e$[N,N]$  result. 

The advantage of the Pad\'e method is that 
it replaces the matrix inversion by a fast estimate of the Born expansion, 
where all terms are evaluated as a matrix-vector multiplication.
This multiplication can be performed as $n$ dot-products of 
two vectors of  $n$ dimension. 
Therefore the calculation requires 
memory to allocate only $2n$, instead of $n^2$,
complex numbers.
This  reduces substantially 
the dimension of the problem.
The price to pay is the recalculation of the matrix lines.

As we will see in the following section, 11 to 15 Pad\'e terms are needed for
convergence. The full calculation takes 1h to 2h  
CPU time in a Pentium IV at 3GHz.
The memory size required is 50 MB.

\section{Results and Discussion}
\label{sec4}

In this section we discuss the results obtained for the on- and 
off-mass-shell amplitudes, the fit to the np differential cross section,
and the convergence of the amplitude as a function of the energy.

The numerical results were
checked to satisfy the Optical Theorem:
\be
Im[ {\cal M}_{\lambda_1 \lambda_2,
\lambda_1 \lambda_2}^{++,++}
(\overline{\mbox{p}},1; \overline{\mbox{p}},1)]=
-\frac{m^2 \overline{\mbox{p}}}{4W}
\sum\limits_{\lambda_3 \lambda_4 }
\int_{-1}^1 \frac{dv}{2 \pi} 
|{\cal M}_{\lambda_3 \lambda_4, \lambda_1 \lambda_2,}^{++,++}
(\overline{\mbox{p}},v; \overline{\mbox{p}},1)|^2.
\label{opticoNN}
\ee
Since the optical theorem only probes the 
on-mass-shell amplitudes,
the results
for the off-mass-shell amplitudes were tested
by numerically checking  that they satisfy the symmetry properties as described
in Section IIIB.

\subsection{On-mass-shell Amplitudes and Differential Cross Section}

Asymptotically the state of the two nucleons is
characterized by the individual isospin states. 
Therefore, for the np system, we must  consider  
\be
T^{np}_{ \lambda_1^\prime \lambda_2^\prime, 
\lambda_1 \lambda_2}(\mbox{p}^\prime, u; \mbox{p}, 1)=
\frac{1}{2}
T^{10}_{ \lambda_1^\prime \lambda_2^\prime, 
\lambda_1 \lambda_2}(\mbox{p}^\prime, u; \mbox{p}, 1)+
\frac{1}{2}
T^{00}_{ \lambda_1^\prime \lambda_2^\prime, 
\lambda_1 \lambda_2}(\mbox{p}^\prime, u; \mbox{p}, 1),
\ee
where 
$T^{I0}_{ \lambda_1^\prime \lambda_2^\prime, 
\lambda_1 \lambda_2}$ represents the anti-symmetrized 
matrix ${\cal M}$ with a total isospin $I$ ($I_z=0$).

Figs.\ \ref{ampM1} 
and \ref{ampM2} show the scattering amplitudes 
(both real and imaginary parts)
obtained, 
at a fixed energy $T_{lab}=300$ MeV, 
for all independent helicity channels. The comparison between the exact result
and the PWD results with increasing values of
the total angular momentum $J$ is shown. 

The convergence of the  Pad\'e amplitudes
has been carefully examined, and the 
iteration procedure stopped when both real and imaginary 
parts of $T$ converge with a relative error lower than $10^{-2}$.
We conclude that in general grids with $N_p=20$ $N_u=30$ were enough for the 
numerical convergence of the solution, although
$N_p=N_u=20$ were sufficient below 200 MeV.

For the helicity $\lambda_1 \lambda_2 =+-$ channel convergence requires 
13 Pad\'e terms for 
$I=0$ and 11 terms for $I=1$. For the $\lambda_1 \lambda_2 =++$ channel
15 terms for $I=0$ and 13 terms for $I=1$ are necessary. 
In all cases optical theorem has always satisfied with 
a relative error smaller than $10^{-2}$.

After calculating all the on-mass-shell polarized amplitudes we can evaluate 
the differential cross section
\be
\frac{d \sigma}{d \Omega} (\overline{\mbox{p}},u)=
\frac{1}{(2\pi)^2} \frac{m^4}{W^2} \overline{|T^{np}
(\overline{\mbox{p}},u) |^2}. 
\ee
In this equation $\overline{|T^{np}|^2}$ is given by 
\be
\overline{|T^{np} 
(\overline{\mbox{p}},u)|^2} 
=\frac{1}{4}\sum\limits_{{
\matrix{ {\lambda_1^\prime,\lambda_2^\prime}\cr 
{\lambda_1,\lambda_2}\cr}} } 
|T^{np}_{\lambda_1^\prime \lambda_2^\prime, \lambda_1 \lambda_2} 
(\overline{\mbox{p}},u; \overline{\mbox{p}},1)|^2, 
\ee
which corresponds to an average over the initial helicity states and 
a sum of the final ones. 
In Fig.\  \ref{figPWD}  the fit of the NN potential model to the 
differential cross section np data at energies $T_{lab}$=99, 
200 and 319 MeV 
is shown.
Data are collected from the Nijmegen data-basis \cite{NNonline} 
and correspond to Refs.\  \cite{Scanlon63,Hurster90,Keeler82}.   

At least for the two first energies the quality of the fit is very good. 
The $T_{lab}=319$ MeV energy case is already very near 
the pion production threshold,
which justifies the slight decrease of the quality of the fit. Nevertheless,
we conclude from this fit that the method developed in the present work is reliable and promising.
Another point worth mentioning is that the fit selects the 
PV pion-nucleon coupling (mixing parameter 
$\lambda_\pi=0$), in
agreement with requirements from chiral symmetry.

\subsection{Off-mass-shell Amplitudes}

The off-mass-shell 
$T^{np}_{ \lambda_1^\prime \lambda_2^\prime, 
\lambda_1 \lambda_2}(\mbox{p}^\prime, u; 
\overline{\mbox{p}}, 1)$ amplitudes may be plotted 
as functions of 2 variables in momentum space.

The results for the 8 $\rho^\prime=+1$ amplitudes are presented in 
Figs.\ \ref{folha1+} and \ref{folha2+} for 
$T_{lab}=300$ MeV.
The $\rho^\prime=-1$ amplitudes plots 
are presented in Figs.\  \ref{folha1-} and \ref{folha2-}.
Note that the on-mass-shell region corresponds to the line
$\mbox{p}^\prime=$0.375 GeV.

It is interesting to note that the off-mass-shell 
region is very important for all channels. 
In fact the magnitudes of some amplitudes are important up to 
momenta of the order 1.5 GeV, much larger than the 
on-mass-shell momentum. It is also important to point out 
that the amplitudes involving transitions to negative-energy states, 
$\rho^\prime =-1$, have magnitudes of the same order of 
the $\rho^\prime=+1$ amplitudes. Therefore, degrees of freedom 
corresponding to negative-energy states may have
considerable weight within covariant effective 
field theories. Consequently, although explicitly absent
in non relativistic models, they are
accounted  for in an effective way through the fitting
procedures yielding
low energy non relativistic realistic potentials.

\section{Breaking of PWD for high energies}

\begin{figure}
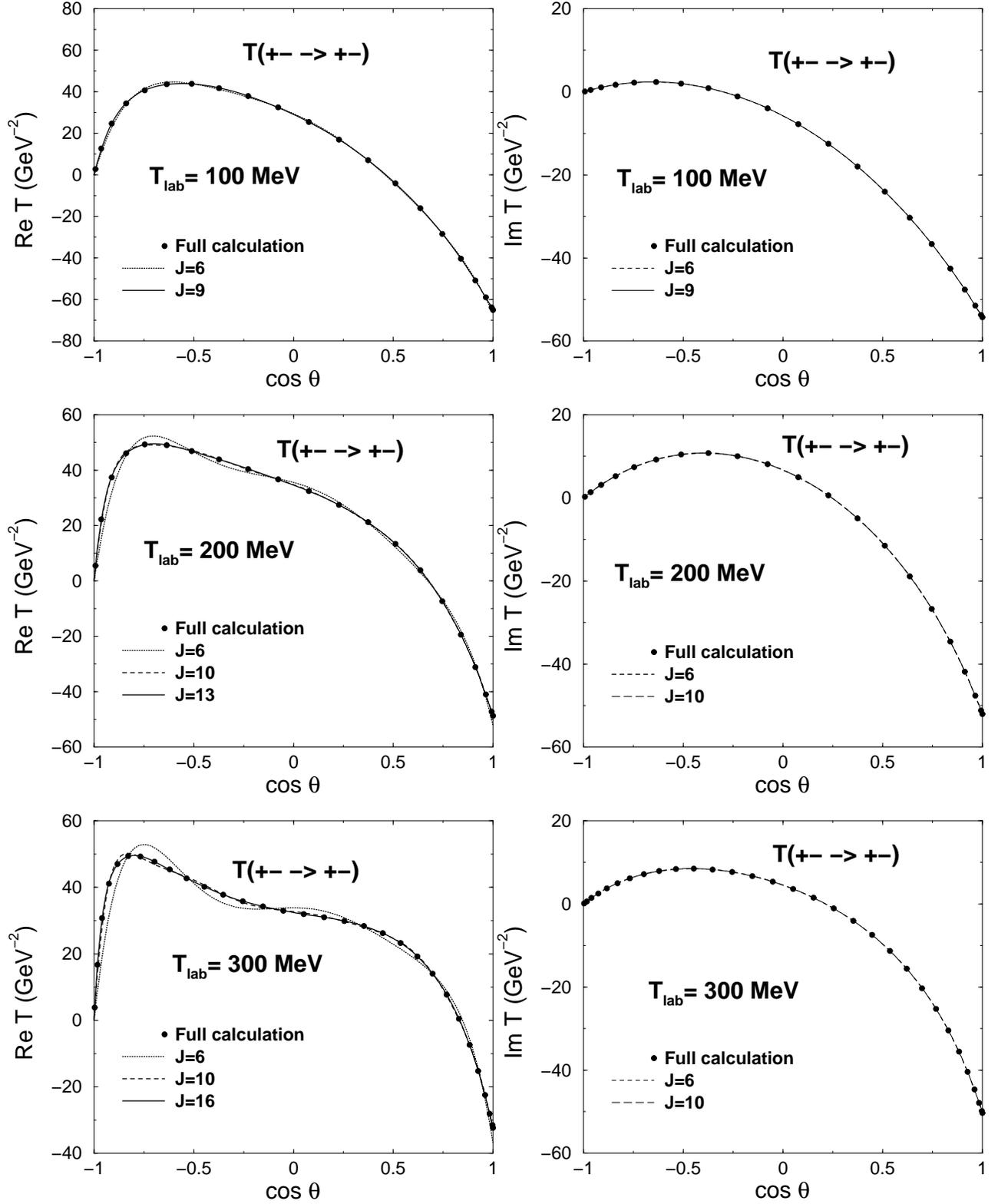

\caption{
Partial wave decomposition of the $M_3$ amplitude 
for 100, 200 and 300 MeV.} 
\label{figPWD}
\end{figure}

We present here the study of the PWD convergence as a function of the energy.
To perform this decomposition 
we follow Ref.\  \cite{BrownJackson}. 
Figure \ref{figPWD} 
shows the convergence of the PWD to the full calculation
for a particular neutron-proton on-mass-shell $T$ matrix, 
for three energy cases.
We notice here that the imaginary part of $T$ converges 
faster than the real part. 
For each energy case the criterium for convergence was defined
as a deviation less than 1\% from the full result.

We can see in the Fig.\  \ref{figPWD} that for 200 MeV 
more that 10 partial waves are needed to achieve convergence. 
We confirmed that some helicity 
channels at $300$ MeV require 16 partial waves,
as obtained in the non-relativistic calculation of Ref.\  \cite{Fac2}. 
As mentioned before,  
this large number of partial waves makes the PWD
method impracticable to generate the two-body amplitudes 
required as an input for the three-body
bound or scattering calculation 
as seen in Ref.\  \cite{Fac3} within 
a non relativistic framework.

On the other hand, the results for $100$ MeV show that 
it is a good approximation at low energies to take $J=6$  as the highest total angular momentum.
This explicit finding fully justifies the cut-off at
$J=6$  reported in the three-body bound state relativistic calculations of 
Ref.\  \cite{Stadler97}.

\section{Conclusions}
\label{sec5}

We considered the Spectator 
equation for nucleons. Its solution depends 
on the helicities of the particles, as well as on their $\rho$-spins. 
($\rho^\prime =+1$ for the positive- energy 
state component and $\rho^\prime =-1$ for 
the negative-energy state component).

The scattering amplitude is solved without partial 
wave decomposition using the Pad\'e method.
This method revealed to be efficient and suitable for the solution 
of the relativistic integral equation without partial wave decomposition.
For $T_{lab}<$ 350 MeV, from 11 to 15 Pad\'e terms  were needed for
convergence.   

Strong form factors 
and a prescription for the exchange kernel 
different from the ones considered in other calculations
within the spectator formalism are used.
When both particles are on-mass-shell 
the present prescription for the exchange kernel
gives rise to the kernel directly 
obtained from the Feynman rules.
This is important in view of 
possible applications to problems where the 
fields are not effective, the $q \bar q$ bound state 
and the quark exchange diagrams in $\pi \pi$ scattering.

We fitted an OBE interaction 
to the differential np differential cross section  
in  the 100-350 MeV energy range.
We let the percentage of
PS and PV admixture to float as a free parameter of the fit. 
It turned
out that the PV coupling was favored by the fit,
which is in agreement arguments of chiral symmetry.

We studied the convergence of the PWD method as a function of the energy.
We concluded that above $T_{lab}=250$ MeV at least 16 partial waves have to be
included for some helicity cases. This large number of terms indicates 
the breaking of the PWD method for applications in heavier nuclei.

Beyond the $\rho^\prime =+1$ amplitudes 
involved in the cross section calculation, we have also calculated the 
$\rho^\prime =-1$ 
amplitudes related to processes involving one off-mass-shell particle. 
These amplitudes are numerically significant and may be used in the calculation of 
meson production cross sections \cite{Lee93,Kolk01} and of the 
$^3$He$(e,e^\prime)$X reaction 
observables for 1-10 GeV energies \cite{CLASS03}.

\bigskip
\begin{center}
{\bf ACKNOWLEDGMENTS}
\end{center}

This work was performed partially under the grant PRAXIS XXI BD/9450/96
and grant POCT/FNU/50358/2002.
One of the authors (G. R.) would like to thank 
Alfred Stadler for helpful discussions.
M. T. P. and G. R. thank Franz Gross and 
Charlotte Elster for discussions, and 
the Jefferson Laboratory Theory group for the
hospitality during their visit.

\appendix

\section{States of Helicities}
\label{apHelicidades}

\subsection{Positive-energy state spinors}

Following the construction of Refs.\  \cite{JacobWick,Wick}
we obtain for the spinors the expressions:
\be
u_j({\bf k},\lambda)=
 N_k 
\left[
\matrix{ {1}\cr {\lambda \tilde k}\cr }
\right] 
|\lambda>_j,
\ee
with the normalization 
\ba
N_k=\sqrt{\frac{m+E_{\bf k}}{2m}} \\
\tilde k= \frac{\mbox{k}}{m+E_{\bf k}}.
\ea

The  Pauli spinors of the  particle 1 and 2
\ba
& & |\lambda>_1 = \chi_{\lambda} ({\bf \hat k}) \\
& &|\lambda>_2 = \psi_{\lambda} ({\bf \hat k}).
\ea
are related by
\be
\psi_\lambda ({\bf \hat k})=
\chi_{-\lambda} ({\bf \hat k}).
\ee
Initial and final state Pauli spinors 
are presented in table \ref{tabHel}.

\subsection{Negative-energy spinor states}

The negative-energy spinors are constructed from positive-energy states 
using the charge conjugation operator $C$ \cite{Gross92,JacobWick,Wick}:
\ba
& &
v_1({\bf k},\lambda)= (-1) \lambda 
C \bar u_2^T({\bf k},\lambda) \\
& &
v_2({\bf k},\lambda)=  \lambda  
C \bar u_1^T({\bf k},\lambda),
\ea
where $C=-i \gamma^0 \gamma^2$ 
and $T$ indicates matrix transposition. 
The relative factors are introduced by 
convenience.
As a result  we get
\be
v_i({\bf k,\lambda})=
N_k
\left[
\matrix{ {-\lambda \tilde k}\cr {1}\cr }
\right]
|\lambda>_i.
\ee

\section{Kernel $\bar {\cal V}$}
\label{apKernel}

In this Appendix we describe the analytical structure of the kernel
with the OBE form.
As mentioned before, for any isospin state $I$, the kernel 
$\bar {\cal V}$ contains the direct, ${\cal V}$, term and the 
exchange kernel, $\widehat {\cal V}$, term:
\be
\bar {\cal V}_{\lambda_1^\prime \lambda_2^\prime, 
\lambda_1 \lambda_2}^{+\,\rho^\prime,+\,\rho}(p^\prime,k;P)=
{\cal V}_{\lambda_1^\prime \lambda_2^\prime, 
\lambda_1 \lambda_2}^{+\,\rho^\prime,+\,\rho}(p^\prime,k;P)
+(-1)^I
\widehat{\cal V}_{\lambda_1^\prime \lambda_2^\prime, 
\lambda_1 \lambda_2}^{+\,\rho^\prime,+\,\rho}(p^\prime,k;P).
\ee

For meson $i$ the direct term is given by
\ba
{\cal V}_{\lambda_1^\prime\lambda_2^\prime, 
\lambda_1 \lambda_2}^{+\;\rho^\prime,+\;\rho}(p^\prime,k;P)&=& 
\delta_I \frac{g_i^2}{\mu_i^2-q^2}
\bar u^+(p_1^\prime,\lambda_1^\prime) 
\Lambda_1(p_1^\prime,k_1)u^+(k_1,\lambda_1) \label{kernelOBEb}
\\
& &
\bar u^{\rho^\prime}(p_2^\prime,\lambda_2^\prime) 
\Lambda_2(p^\prime_2,k_2)u^{\rho}(k_2,\lambda_2)
[f_{m_i}(q^2)]^2 f_N(p_2^{\prime\;2})f_N(k_2^2).
\nonumber 
\ea
where the exchange momentum is
\be
q^2=(E_{{\bf p}^\prime}-E_{\bf k})^2-
({\bf p}^\prime-{\bf k})^2.
\ee

For vector mesons we have to undertake the substitution
\be
\Lambda_1 \Lambda_2 \to
\Lambda_1^\mu (p_1^\prime,k_1)
\Lambda_2^\nu (p_2^\prime,k_2)
\left[g_{\mu \nu} +
\frac{(p_1^\prime-k_1)_\mu (p_2^\prime-k_2)_\nu} {\mu_i^2}
\right].
\label{propagaV}
\ee

Similarly for the exchange term we have
\ba
\widehat
{\cal V}_{\lambda_1^\prime\lambda_2^\prime, 
\lambda_1 \lambda_2}^{+\;\rho^\prime,+\;\rho}(p^\prime,k;P)
&=& 
\delta_I \frac{g_i^2}{\mu_i^2-\widehat q^2}
\bar u^+(p_1^\prime,\lambda_1^\prime) 
\Lambda_1(p_1^\prime,k_2)u^{\rho}(k_2,\lambda_2)
\label{kernelOBE2b} \\
& &
\bar u^{\rho^\prime}(p_2^\prime,\lambda_2^\prime) 
\Lambda_2(p^\prime_2,k_1)u^+(k_1,\lambda_1)
[f_{m_i}(\widehat q^2)]^2 f_N(p_2^{\prime\;2})f_N(k_2^2),
\nonumber 
\ea
where, according to the "on-mass-shell prescription" the exchange momentum is
\be
\widehat q^2
=(E_{{\bf p}^\prime}-E_{\bf k})^2-
({\bf p}^\prime+{\bf k})^2.
\ee

For vector mesons the replacement is now 
\be
\Lambda_1 \Lambda_2 \to
\Lambda_1^\mu (p_1^\prime,k_2)
\Lambda_2^\nu (p_2^\prime,k_1) \left[g_{\mu \nu} +
\frac{(p_1^\prime-k_2)_\mu (p_2^\prime-k_1)_\nu} {\mu_i^2}
\right].
\label{propagaV2}
\ee

The calculation results are presented in Sec.\  \ref{kdirecto} for the 
direct term and Sec.\  \ref{ktroca} for the exchange 
term. 
For the sake of simplicity 
we do not explicit the parameters involved 
but only the analytical structure.

\subsection{Direct Kernel}
\label{kdirecto}

We will write the direct kernel 
by means of the following auxiliary functions 
\ba
& &
Z_1^0({\bf \hat p^\prime},{\bf \hat k})=
\chi_{\lambda_1^\prime}^{\prime\; \dagger} ({\bf \hat p^\prime})
\chi_{\lambda_1}({\bf \hat k})  
\label{Z10}\\
& &
Z_2^0({\bf \hat p^\prime},{\bf \hat k})=
\psi_{\lambda_2^\prime}^{\prime\; \dagger} ({\bf \hat p^\prime})
\psi_{\lambda_2}({\bf \hat k}),  
\label{Z20}
\ea
\ba
& &
Z_1^i({\bf \hat p^\prime},{\bf \hat k})=
\chi_{\lambda_1^\prime}^{\prime\; \dagger} ({\bf \hat p^\prime})
\sigma_i^{(1)}  
\chi_{\lambda_1}({\bf \hat k})  
\label{Z1i}\\
& &
Z_2^i({\bf \hat p^\prime},{\bf \hat k})=
\psi_{\lambda_2^\prime}^{\prime\; \dagger} ({\bf \hat p^\prime})
\sigma_i^{(2)}  
\psi_{\lambda_2}({\bf \hat k})  
\label{Z2i}
\ea
The $Z_j^\alpha$ functions ($j=1,2$, $\alpha=0,..,3$) 
are calculated from the Pauli spinors presented in 
Appendix \ref{apHelicidades}.
For simplicity sometimes we suppress the arguments of $Z_j^\alpha$.

\subsubsection{Meson $\sigma$ }

\ba
{\cal V}_{\lambda_1^\prime\lambda_2^\prime, 
\lambda_1 \lambda_2}^{+\;\rho^\prime,+\;\rho}(p^\prime,k)
&=&
-\frac{g_\sigma^2}{\mu_\sigma^2-q^2} [f_\sigma(q^2)]^2 
f_N(p^{\prime\;2})f_N(k^2)
\nonumber \\
& &N_{p^\prime}^2 N_k^2 H_\sigma(p^\prime,k) \\
& &Z_1^0({\bf \hat p^\prime},{\bf \hat k}) Z_2^0({\bf \hat p^\prime},
{\bf \hat k})
\nonumber
\ea

$H_\sigma(p^\prime,k)$ is also a function of 
$\lambda_1^\prime$, $\lambda_2^\prime$, $\lambda_1$, $\lambda_2$, 
$\rho^\prime$ and $\rho$.

For large $\mbox{k}$ we have $H_{\sigma} (p^\prime,k)  \sim 1$.

\subsubsection{Meson $\pi$}

\ba
{\cal V}_{\lambda_1^\prime\lambda_2^\prime, 
\lambda_1 \lambda_2}^{+\;\rho^\prime,+\;\rho}(p^\prime,k;P)
\label{Vpi}
&=&
\delta_I
\frac{g_\pi^2}{\mu_\pi^2-q^2} [f_\pi(q^2)]^2 
f_N(p^{\prime\;2})f_N(k^2)
\nonumber \\
& &N_{p^\prime}^2 N_k^2 H_{\pi}(p^\prime,k) \\
& &Z_1^0({\bf \hat p^\prime},{\bf \hat k}) Z_2^0({\bf \hat p^\prime},
{\bf \hat k})
\nonumber
\ea

$H_\pi(p^\prime,k)$ is also a function of $\lambda_1^\prime$, $\lambda_2^\prime$, 
$\lambda_1$, $\lambda_2$, $\rho^\prime$ and $\rho$. 

For large $\mbox{k}$ we have $H_\pi (p^\prime,k) \sim 1$ for PS coupling and 
$H_\pi (p^\prime,k) \sim \mbox{k}/m$ for PV coupling.

\subsubsection{Vector mesons}
 
\ba
{\cal V}_{\lambda_1^\prime\lambda_2^\prime, 
\lambda_1 \lambda_2}^{+\;\rho^\prime,+\;\rho}(p^\prime,k;P)
&=&
\delta_I
\frac{g_v^2}{\mu_v^2-q^2} [f_v(q^2)]^2  
f_N(p^{\prime\;2})f_N(k^2)
\nonumber \\
& &N_{p^\prime}^2 N_k^2 
H_v (p^\prime,k).
\label{eqVvecDir}
\ea

\ba
H_v(p^\prime,k)
&=&
r_0Z_1^0 Z_2^0 
+r_1 \sum\limits_{i=1}^3 Z_1^i Z_2^i  
+r_2 \sum\limits_{i=1}^3(p_i^\prime + k_i) Z_1^0 Z_2^i
\nonumber \\
& &
+r_3\sum\limits_{i=1}^3(p_i^\prime + k_i) Z_1^i Z_2^0
+r_4\sum\limits_{i=1}^3(p_i^\prime + k_i)^2 Z_1^0 Z_2^0.
\label{eqHv} 
\ea

The coefficients $r_l$ ($l=0,..,4$) are functions of the momenta 
$\mbox{p}^\prime$, $\mbox{k}$, of the indices 
$\lambda_1^\prime$, $\lambda_2^\prime$, 
$\lambda_1$, $\lambda_2$, $\rho^\prime$, $\rho$ and of the 
$\kappa_v$ parameter.

For large $\mbox{k}$ we have $H_v (p^\prime,k) \sim 1$ when $\kappa_v =0$ 
and $H_v (k,k) \sim \mbox{k}^2/m^2$ and $H_v (p^\prime,k) \sim \mbox{k}/m$ 
when $\kappa_v \ne 0$.

\subsection{Exchange Kernel}
\label{ktroca}

For the exchange term we use the auxiliary functions
\ba
& &
\widehat Z_1^0({\bf \hat p^\prime},{\bf \hat k})=
\chi_{\lambda_1^\prime}^{\prime\; \dagger} ({\bf \hat p^\prime})
\psi_{\lambda_2}({\bf \hat k})  
\label{hatZ10}\\
& &
\widehat Z_2^0({\bf \hat p^\prime},{\bf \hat k})=
\psi_{\lambda_1^\prime}^{\prime\; \dagger} ({\bf \hat p^\prime})
\chi_{\lambda_1}({\bf \hat k}), \label{hatZ20}
\ea
\ba
& &
\widehat Z_1^i({\bf \hat p^\prime},{\bf \hat k})=
\chi_{\lambda_1^\prime}^{\prime\; \dagger} ({\bf \hat p^\prime})
\sigma_i^{(1)}  
\psi_{\lambda_2}({\bf \hat k})  
\label{hatZ1i}\\
& &
\widehat Z_2^i({\bf \hat p^\prime},{\bf \hat k})=
\psi_{\lambda_2^\prime}^{\prime\; \dagger} ({\bf \hat p^\prime})
\sigma_i^{(2)}  
\chi_{\lambda_1}({\bf \hat k}).
\label{hatZ2i}
\ea
These functions are also evaluated from the Pauli spinors.

\subsubsection{Meson $\sigma$ }

\ba
\widehat {\cal V}_{\lambda_1^\prime\lambda_2^\prime, 
\lambda_1 \lambda_2}^{+\rho^\prime,+\;\rho}(p^\prime,k;P)
&=&
-\frac{g_\sigma^2}{\mu_\sigma^2-{\widehat q}^2} 
[f_\sigma({\widehat q}^2)]^2 f_N(p^{\prime\;2})f_N(k^2)
\nonumber \\
& &N_{p^\prime}^2 N_k^2 \widehat H_{\sigma}(p^\prime,k) \\
& &\widehat Z_1^0({\bf \hat p^\prime},{\bf \hat k}) 
\widehat Z_2^0({\bf \hat p^\prime},{\bf \hat k})
\nonumber
\ea

$\widehat H_\sigma(p^\prime,k)$ is also a function 
of $\lambda_1^\prime$, $\lambda_2^\prime$, 
$\lambda_1$, $\lambda_2$, $\rho^\prime$ and $\rho$.

For large $\mbox{k}$ we have $\widehat H_{\sigma} (p^\prime,k) \sim 1$.

\subsubsection{Meson $\pi$ }

\ba
\widehat{\cal V}_{\lambda_1^\prime\lambda_2^\prime, 
\lambda_1 \lambda_2}^{+\rho^\prime,+\rho}(p^\prime,k;P)
\label{Vpi2}
&=&
\delta_I
\frac{g_\pi^2}{\mu_\pi^2-q^2} [f_\pi(q^2)]^2 f_N(p^{\prime\;2})f_N(k^2)
\nonumber \\
& &N_{p^\prime}^2 N_k^2 \widehat H_{\pi}(p^\prime,k) \\
& &\widehat Z_1^0({\bf \hat p^\prime},{\bf \hat k}) 
\widehat Z_2^0({\bf \hat p^\prime},{\bf \hat k})
\nonumber
\ea

$\widehat H_{\pi} (p^\prime,k)$ is also a function  of
$\lambda_1^\prime$, $\lambda_2^\prime$, 
$\lambda_1$, $\lambda_2$, $\rho^\prime$ and $\rho$. 

For large $\mbox{k}$ we have $\widehat H_\pi (p^\prime,k) \sim 1$ 
for PS coupling and $\widehat H_\pi (p^\prime,k) \sim \mbox{k}/m$ 
and $\widehat H_\pi (k,k) \sim \mbox{k}^2/m^2$ for PV coupling.

\subsubsection{Vector mesons}

\ba
\widehat {\cal V}_{\lambda_1^\prime\lambda_2^\prime, 
\lambda_1 \lambda_2}^{+\;\rho^\prime,+\;\rho}(p^\prime,k;P)
&=&
\delta_I
\frac{g_v^2}{\mu_v^2-q^2} [F_v(q^2)]^2 \nonumber \\
& &N_{p^\prime}^2 N_k^2 \widehat H_v (p^\prime,k).
\label{eqVvecTroca}
\ea

\ba
\widehat H_v(p^\prime,k)
&=&
r_0 \widehat Z_1^0 \widehat Z_2^0 
+r_1 \sum\limits_{i=1}^3 \widehat Z_1^i \widehat Z_2^i 
+r_2 \sum\limits_{i=1}^3(p_i^\prime - k_i) \widehat Z_1^0 \widehat Z_2^i
\nonumber  \\
& &
+r_3\sum\limits_{i=1}^3(p_i^\prime - k_i) \widehat Z_1^i \widehat Z_2^0
+r_4\sum\limits_{i=1}^3(p_i^\prime - k_i)^2 \widehat Z_1^0 \widehat Z_2^0.
\label{eqHv2} 
\ea

The coefficients $r_l$ ($l=0,..,4$) are a function the momenta 
$\mbox{p}^\prime$, $\mbox{k}$, of $\lambda_1^\prime$, $\lambda_2^\prime$, 
$\lambda_1$, $\lambda_2$, $\rho^\prime$, $\rho$ and also of the 
$\kappa_v$ parameter. The $r_l$ coefficients for the exchange kernel 
are not related with the direct kernel coefficients.

For large $\mbox{k}$ we have $\widehat H_v (p^\prime,k) \sim 1$ 
when $\kappa_v =0$ and $\widehat H_v (k,k) \sim \mbox{k}^2/m^2$ and 
$\widehat H_v (p^\prime,k) \sim \mbox{k}/m$ when   
$\kappa_v \ne 0$.

\section{$\varphi$ rotated amplitude}
\label{apAmpRot}

In this appendix we derive the relation between 
the scattering amplitude on the scattering plane
(${\bf p}$ is on the $z$ axis)  

$$
{\cal M}_{\lambda_1^\prime \lambda_2^\prime,
\lambda_1 \lambda_2}^{+ \;\rho_2^\prime,+ \;\rho_2}
(\mbox{p}^\prime,\theta^\prime,0; \mbox{p};W)
$$
and the scattering amplitude on a rotated
plane characterized by  $\varphi^\prime$-rotation   
in the $z$ axis
$$
{\cal M}_{\lambda_1^\prime \lambda_2^\prime,
\lambda_1 \lambda_2}^{+ \;\rho_2^\prime,+ \;\rho_2}
(\mbox{p}^\prime,\theta^\prime,\varphi^\prime; \mbox{p};W).
$$

Thus, we consider the Lorentz transformation 
\be
\Lambda = R_{-\varphi^\prime,0,0}
\label{rodaPhi}
\ee
corresponding to a rotation of an angle $\varphi^\prime$
around the $z$ axis.

The correspondence 
between the spinors before and after this Lorentz 
transformation is
\ba
& &S(\Lambda) u_1^\rho(p,\lambda)= \sum\limits_{\lambda^\prime} 
D_{\lambda^\prime \lambda} (R_\Lambda) 
u_1^\rho(\Lambda p, \lambda^\prime) \label{trans1}\\
& &S(\Lambda) u_2^\rho(p,\lambda)= \sum\limits_{\lambda^\prime} 
D_{-\lambda^\prime,-\lambda} (R_\Lambda) 
u_2^\rho(\Lambda p, \lambda^\prime), \label{trans12}
\ea
where $S(\Lambda)$ is the operator that 
transforms the $u,v$-states in the $\Lambda$ 
Lorentz transformation and 
$D$ is the usual $D^{1/2}$ Wigner matrix
in terms of the rotation angles.
The rotation operators are given by 
\ba
& & R_\Lambda= H_{\Lambda p}^{-1} \Lambda H_{p} 
\label{RLdef}\\
& & R_\Lambda^\prime= H_{\Lambda{p^\prime}}^{-1} 
\Lambda H_{p^\prime}. 
\label{RLpdef}
\ea
In the last equations 
$H_p$ is the operator that transforms 
a 4-momentum $(m,{\bf 0})$ into $p=(E_{\bf p},{\bf p})$.
Details can be found in Ref.\  \cite{Carruthers}. 
The operation can always be written in a sequence of 
a boost ($L_p$) and a rotation ($R_{\hat {\bf p}}$)  
\be
H_p= R_{\hat {\bf p}} L_p.
\ee

After the Lorentz transformation has been done 
the following relation between the original and the rotated 
scattering amplitudes is obtained
\ba
& &
{\cal M}_{\alpha^\prime \beta^\prime, \alpha \beta}
(\Lambda p^\prime,\Lambda p; \Lambda P)= 
\label{eqTLdirac}
\\
& &
\sum\limits_{{
\matrix {{\alpha_1, \beta_1}\cr {\alpha_2, \beta_2}\cr}}}
S_{\alpha^\prime \alpha_1}(\Lambda) S_{\beta^\prime \beta_1}(\Lambda) 
{\cal M}_{\alpha_1 \beta_1, \alpha_2 \beta_2}(p^\prime,p;P)
S_{\alpha_2 \alpha}^{-1}(\Lambda) S_{\beta_2 \beta}^{-1}(\Lambda). 
\nonumber
\ea
for more details see and Appendix B of Ref.\ \cite{Gross92}.  

Changing (\ref{eqTLdirac}) 
to the helicity representation according to 
eq.\  (\ref{Mlambda}) and using the operator rotation 
properties and the invariance of permutation 
between boost and rotation operators 
related to the same axis, we finally 
conclude  that
\be
{\cal M}_{\lambda_1^\prime \lambda_2^\prime,
\lambda_1 \lambda_2}^{+ \;\rho_2^\prime,+ \;\rho_2}
(\mbox{p}^\prime,\theta^\prime,\varphi^\prime; \mbox{p};W)=
e^{i\frac{\lambda_1-\lambda_2}{2}\varphi^\prime}
{\cal M}_{\lambda_1^\prime \lambda_2^\prime,
\lambda_1 \lambda_2}^{+ \;\rho_2^\prime,+ \;\rho_2}
(\mbox{p}^\prime,\theta^\prime,0; \mbox{p};W).
\label{eqTransMb}
\ee

\section{Function $V$}
\label{apIntPhi}

In this Appendix we describe 
how to evaluate $V$ defined by Eq.\ (\ref{defV}).
By performing the $\varphi$ integration we get:
\be
V(\mbox{p}^\prime,\theta^\prime,\mbox{k},\theta;\bar \lambda,W)= 
\int_0^{2\pi}
e^{i {\bar \lambda} \varphi}
\bar {\cal V}(\varphi) d \varphi,
\label{intVphi}
\ee
where we use the simplification
\be
\bar {\cal V}(\varphi) =
\bar {\cal V}_{\lambda_1^\prime \lambda_2^\prime, 
\lambda_3 \lambda_4}^{+ \rho_2^\prime, + \rho}
(\mbox{p}^\prime,\theta^\prime,0;
\mbox{k},\theta,\varphi;W).
\ee

We note  that
\be
\bar \lambda=\frac{\lambda_1-\lambda_2}{2},
\ee
so $\bar \lambda=0,\pm 1$.

First,  we separate the $\varphi$-dependent parts 
from the independent ones (factorization). 
Next, we analyze 
the structure of the resulting functions.
We need to consider two different cases: the 
non vector meson exchange ($\sigma$ and  $\pi$) 
and the vector meson exchange.

\subsection{Factorization of $\bar {\cal V}(\varphi)$}

From the ${\cal V}$ expression 
(see Appendix \ref{apKernel}) we conclude 
that 
\be
\bar 
{\cal V}(\varphi)=\delta_I \Lambda_m^4 g_m^2 
\bar V(\varphi) 
{\cal R},
\label{eqVbarPHI}
\ee
where
\be
\bar V(\varphi) =
\frac{[f_{m_i}(q^2)]^2}
{\Lambda_m^4}
\frac{g_i^2}{\mu^2-q^2}, 
\ee
and
\ba
{\cal R}=
N_{p^\prime}^2 N_{k}^2
f_N(p^{\prime\; 2}) f_N(k^2) 
\tilde H_i (p^\prime,k).
\label{eqR1}
\ea
In this equations $i$ is the meson index. 
Also from Appendix \ref{apKernel} we have
\be
\tilde H_i (p^\prime,k)= H_i (p^\prime,k) Z_1^0 Z_2^0,
\ee
for a non vector meson and
\be
\tilde H_i (p^\prime,k)= H_i (p^\prime,k),
\ee
for vector mesons.
Note that the parameterization (\ref{eqR1}) 
is valid for the direct term and for the 
exchange term if we replace $q^2$ by $\widehat q^2$,  
$H_i$ by $\widehat H_i$ 
(and $Z_j^0$ by $\widehat Z_j^0$ for the non vector case).

Using the coordinates definition 
(\ref{eqCoor1})-(\ref{eqCoor3})
with $\varphi^\prime=0$ for $q^2$ and 
$\widehat q^2$ we can conclude that
\be
\bar V(\varphi) =
\frac{1}{a-b \cos \varphi}
\frac{1}{(c-b \cos \varphi)^2},
\label{eqVphi0}
\ee
where
\ba
& &
a=\mu^2+\mbox{p}^{\prime\; 2}+\mbox{k}^2 
\mp 2\mbox{p}^\prime \mbox{k} 
\cos \theta^\prime \cos \theta -q_0^2\\
& &
b=\pm 2 \mbox{p}^\prime \mbox{k} 
\sin \theta^\prime \sin \theta. \\
& &
c=\Lambda_m^2+\mbox{p}^{\prime\; 2}+\mbox{k}^2 \mp
2\mbox{p}^\prime \mbox{k} \cos \theta^\prime \cos \theta
-q_0^{2}.
\ea
The upper sign should be used in the  
direct term and the lower sign in the exchange term.

\subsection{${\cal R}$ in terms of $\varphi$}

The next step is to write ${\cal R}$ 
of eq.\   (\ref{eqR1}) in terms of $\varphi$.
We need to consider two separate cases:  
the non vector mesons 
and the vector mesons.

\subsubsection{Non vector mesons}
\label{secNVM}

For non vector mesons we can write 
\be
{\cal R}= f(\mbox{p}^\prime,\mbox{k}) \cdot Z_1^0 Z_2^0.
\ee
The exact expression of $f(\mbox{p}^\prime,\mbox{k})$ 
can be  easily deduced from (\ref{eqR1}).
Attending to the 
$Z_1^{0} Z_2^{0}$ dependence in $\varphi$, 
we conclude that
\be
{\cal R}= f(\mbox{p}^\prime,\mbox{k}) \cdot 
(c_0+c_1 \; e^{-i{\lambda_3\varphi}} 
+c_2 \; e^{i {\lambda_3\varphi}}), 
\ee
where $c_l$ ($l=0,..,2$) are known 
coefficients depending on the  
scattering conditions and on the helicities states.
We can write
\ba
V(\mbox{p}^\prime,\theta^\prime,\mbox{k},\theta;\bar \lambda,W)&=&
\delta_I \Lambda_m^4 g_m^2 f(\mbox{p}^\prime,\mbox{k}) \cdot \nonumber \\
& & 
[c_0 {\cal F}_0(\bar \lambda) +c_1 {\cal F}_0(\bar \lambda-\lambda_3)
+ c_2 {\cal F}_0(\bar \lambda+\lambda_3)],
\label{eqVF0}
\ea
where the function ${\cal F}_0(n)$ is defined as
\be
{\cal F}_0(n)= 
\int_0^{2\pi} d \varphi e^{i n \varphi} \; \bar V(\varphi), 
\ee
and $n$ are
\mbox{$n=0,\pm 1,\pm 2$}.

\subsubsection{Vector mesons}

For vector mesons the function 
${\cal R}$ can be written as 
a linear combination of the terms 
$$
Z_1^{\alpha_1} Z_2^{\alpha_2} 
\;\;\mbox{and} \;\;
k_i Z_1^{\alpha_1} Z_2^{\alpha_2},
$$
where $\alpha_1, \alpha_2 =0,..,3$.
The first term is reduced to the non vector 
case discussed in  subsection  \ref{secNVM},
because 
$Z_1^{\alpha_1} Z_2^{\alpha_2}$ can also be written 
as 
$$
c_0+c_1 \; e^{-i{\lambda_3\varphi}} 
+c_2 \; e^{i {\lambda_3\varphi}},$$ 
with appropriated coefficients.
The second term can be decomposed in 3 cases 
considered as follows:\\

\noindent
{\bf Case 1} ($k_1=\mbox{k} \sin \theta \cos \varphi$) \\

In this case we need to integrate factors like
$$
(\mbox{k} \sin \theta) \cos \varphi e^{i n \varphi} 
\bar V(\varphi),
$$
and the corresponding term of $V$,  
which we label $V_1$, is 
\ba
V_1(\mbox{p}^\prime,\theta^\prime,\mbox{k},\theta;\bar \lambda,W)& =&
\delta_I \Lambda_m^4 g_m^2 f(\mbox{p}^\prime,\mbox{k}) (\mbox{k} \sin \theta)  
\cdot \nonumber \\ 
& &[c_0 {\cal F}_1(\bar \lambda) +c_1 {\cal F}_1(\bar \lambda-\lambda_3)
+ c_2 {\cal F}_1(\bar \lambda+\lambda_3)],
\ea
where  
\be
{\cal F}_1(n)= 
\int_0^{2\pi} d \varphi \cos \varphi \;e^{i n \varphi} \bar V(\varphi), 
\ee
with $n=0,\pm 1 \pm 2$. \\

\noindent
{\bf Case 2} ($k_2=\mbox{k} \sin \theta \sin \varphi$)\\

We have also to consider terms like
$$
(\mbox{k} \sin \theta) \sin \varphi e^{i n \varphi} 
\bar V(\varphi),
$$
from which results for the corresponding $V$, 
which we label $V_2$ 
\ba
V_2(\mbox{p}^\prime,\theta^\prime,\mbox{k},\theta;\bar \lambda,W)&=&
\delta_I \Lambda_m^4 g_m^2 f(\mbox{p}^\prime,\mbox{k}) (\mbox{k} \sin \theta) 
 \cdot \nonumber \\ 
& &[c_0 {\cal F}_2(\bar \lambda) +c_1 {\cal F}_2(\bar \lambda-\lambda_3)
+ c_2 {\cal F}_2(\bar \lambda+\lambda_3)].
\ea

The ${\cal F}_2$ function is defined as 
\be
{\cal F}_2(n)= 
\int_0^{2\pi} d \varphi \sin \varphi \;e^{i n \varphi} \bar V(\varphi), 
\ee
with $n=0,\pm 1,\pm 2$.\\

\noindent
{\bf Case 3} ($k_3=\cos \theta$) \\

No new $\varphi$-dependence appears.
This case reduces to the non vector meson case.

\subsection{Functions ${\cal F}_l(n)$}

The functions ${\cal F}_l(n)$ with $l=0,1,2$
can be written in terms of the integrals
$$
R_l=\int_0^{2 \pi} \bar V(\varphi) \cos^l \; \varphi d\varphi,
$$
for $l=0,..,3$.
The $R_l$ integrals are performed analytically with the 
software program {\it Mathematica}  and simplified afterwards.


\newpage

\begin{table}
\begin{center}
\begin{tabular}{|c|c|}
\hline
$m_\pi$ & 138 MeV \\
$G_\pi$ & 13.470 \\
$\lambda_\pi$ & 0.0 \\
$\Lambda_\pi$ & 1190 MeV \\ \hline
$m_\sigma$ & 497 MeV \\
$G_\sigma$ &  3.782 \\ \hline
$m_\rho$ & 770 MeV \\
$G_\rho$ & 0.100 \\ 
$\kappa_\rho$ & 5.644 \\ \hline
$m_\omega$ & 783 MeV \\
$G_\omega$ & 8.100 \\
$\kappa_\omega$ & 0.337 \\ \hline
$\Lambda_m$ & 2400 MeV \\
$\Lambda_N$ & 1783 MeV \\ 
\hline
\end{tabular}
\caption{Model parameters.}
\label{tabNN}
\end{center}
\end{table}

\begin{table} 
\begin{center}
\begin{tabular}{|c|c|c|c|} 
\hline
$I_0$ & $\rho$ & $\lambda_1$ & $\lambda_2$ \\
\hline
1 &    $+$   &     $-$       &      $-$      \\
2 &    $+$   &     $-$       &      $+$      \\
3 &    $+$   &     $+$       &      $-$      \\
4 &    $+$   &     $+$       &      $+$      \\ 
\hline
5 &    $-$   &     $-$       &      $-$      \\
6 &    $-$   &     $-$       &      $+$      \\
7 &    $-$   &     $+$       &      $-$      \\
8 &    $-$   &     $+$       &      $+$      \\ 
\hline
\end{tabular}
\end{center}
\caption{Table defining the indices $I_0$, $I_k$ or $I^{\prime}$.}
\label{tabelaMI} 
\end{table}

\begin{table}
\begin{center}
\begin{tabular}{|ccccc|} \hline \hline
& & Initial state &  &\\
& &  $\lambda =1$ & & $\lambda =-1$ \\
\hline
& & & & \\
$\chi _{\lambda }({\bf \hat z})$ & & 
$\left( \matrix{{1}\cr {0}\cr} \right)$ & &
$\left( \matrix{{0}\cr {1}\cr} \right)$ \\
& & & & \\
$\psi _{\lambda }({\bf \hat z})$ & & 
$\left( \matrix{{0}\cr  {1}\cr} \right)$ & &
$\left( \matrix{{1}\cr  {0}\cr} \right)$ \\
& & & & \\
\hline
& & Final state& &\\
& & $\lambda ^{\prime}=1$ & &$\lambda ^{\prime}=-1$ \\
\hline
& & & & \\ 
$\chi _{\lambda ^{\prime}}^{\prime}(\theta,\varphi)$ & &
$\left( \matrix{
{\cos\frac{\theta }{2}  e^{-i \frac{\varphi}{2}}}\cr 
{\sin\frac{\theta }{2}  e^{ i \frac{\varphi}{2}}}\cr} \right)$ & &
$\left( \matrix{
{-\sin\frac{\theta}{2}  e^{-i \frac{\varphi}{2}}}\cr 
{ \cos\frac{\theta}{2}  e^{ i \frac{\varphi}{2}}}\cr} \right)$ \\
& & & &\\
$\psi _{\lambda ^{\prime}}^{\prime}(\theta,\varphi)$ & &
$\left( \matrix{
{-\sin \frac{\theta}{2} e^{-i \frac{\varphi}{2}}}\cr 
{ \cos \frac{\theta}{2} e^{ i \frac{\varphi}{2}}}\cr} \right)$ & &
$\left( \matrix{
{ \cos\frac{\theta }{2} e^{-i \frac{\varphi}{2}}}\cr 
{ \sin\frac{\theta }{2} e^{ i \frac{\varphi}{2}}}\cr} \right)$ \\
& & & & \\
\hline \hline
\end{tabular}
\caption{Pauli helicity states. The phase differences 
between these spinors and the ones of these Refs.\ [39,40]
are due to the fact that we consider, for convenience, the rotations 
convention of Ref.\ [40] and the phase convention for particle 2 
of Ref.\ [39].}
\label{tabHel}
\end{center}
\end{table}


\newpage

\begin{figure}
\caption{
Helicity amplitudes for 300 MeV 
and partial wave decomposition.} 
\label{ampM1}
\end{figure} 

\begin{figure}
\caption{
Helicity amplitudes for 300 MeV 
and partial wave decomposition.} 
\label{ampM2}
\end{figure} 

\begin{figure}
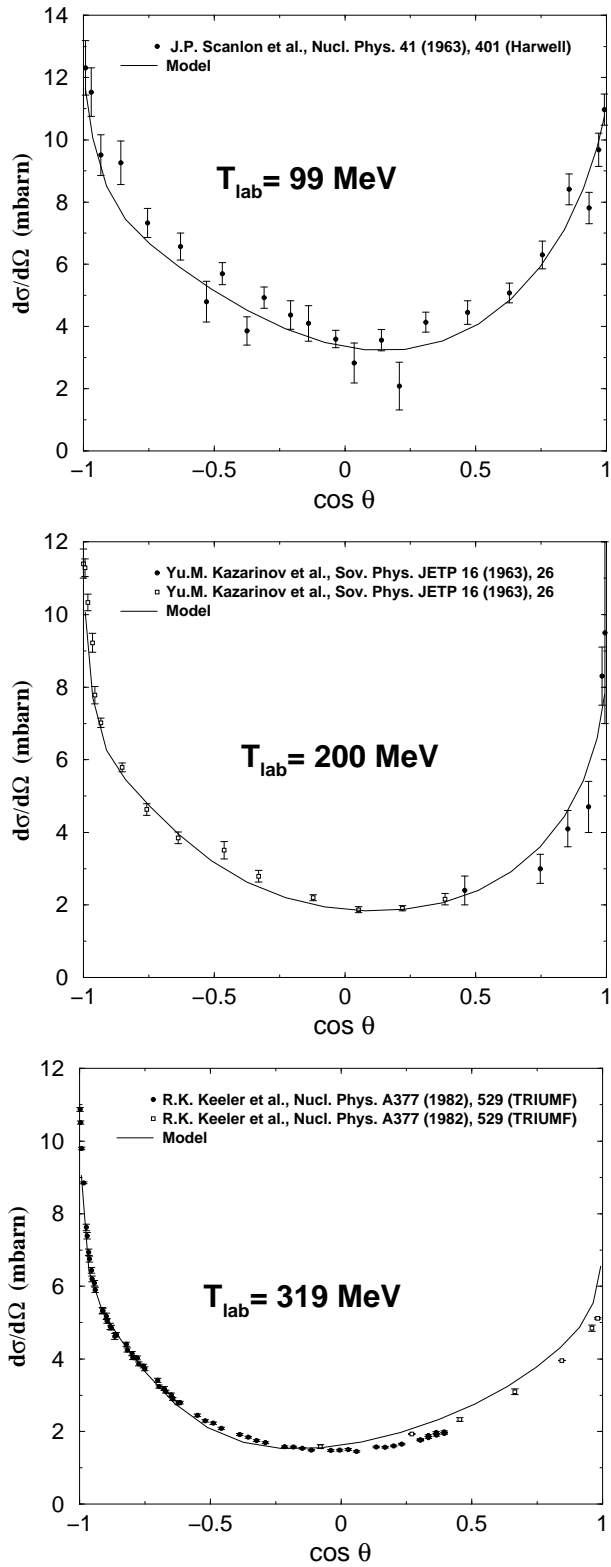

\caption{
Differential cross section results for 99, 200 and 319 MeV.}
\label{DSG}
\end{figure}

\begin{figure}
\caption{
Off-mass-shell amplitudes for 
np process with 
$\rho=+1$, $\rho^\prime=+1$.} 
\label{folha1+}
\end{figure} 

\begin{figure}
\caption{
Off-mass-shell amplitudes for 
np process with 
$\rho=+1$, $\rho^\prime=+1$.} 
\label{folha2+}
\end{figure} 

\begin{figure}
\caption{
Off-mass-shell amplitudes for 
np process with 
$\rho=+1$, $\rho^\prime=-1$.} 
\label{folha1-}
\end{figure} 

\begin{figure}
\caption{
Off-mass-shell amplitudes for 
np process with 
$\rho=+1$, $\rho^\prime=-1$.} 
\label{folha2-}
\end{figure} 

\end{document}